\documentclass[a4paper,12pt]{article}
\usepackage{amssymb}
\usepackage[english]{babel} 
\usepackage{amsfonts}
\usepackage{hyperref}
\usepackage{amsmath}
\usepackage{graphicx}
\usepackage{url}
\usepackage{xcolor}
\usepackage[affil-it]{authblk}
\usepackage{multirow}
\usepackage{etoolbox}
\usepackage[top=2.cm, bottom=2.cm, left=3.5cm, right=3.5cm]{geometry}

\usepackage{textcomp}
\usepackage{amstext}

\date{~}

\begin{document}
\title{NA61/SHINE shining more light on the onset of deconfinement}
\author[]{Maciej P. Lewicki}
\author[]{Ludwik Turko}
\affil[]{on behalf of NA61/SHINE Collaboration}
\maketitle

 \vspace{-1.5cm}
{\it \large \noindent
"It was easier to know it than to explain why I know it." said Sherlock.\\
- Arthur Conan Doyle \footnote{as reminded by J.R. Pelaez in \cite{Pelaez:2015qba}}\\
}

NA61/SHINE has recently completed data acquisition for its original programme on strong interactions. The Collaboration has gathered rich data on collisions of ions in a two-dimensional scan: varying the beam energy and the sizes of colliding nuclei. The most recent analysis of hadron production in $^{40}$Ar+$^{45}$Sc and $^{7}$Be+$^{9}$Be interactions deliver some puzzling results which none of the theoretical models can reproduce.

\section{The NA61/SHINE experiment at CERN}

NA61/SHINE (SPS Heavy Ion and Neutrino Experiment) \cite{facility} is a multi-purpose experiment studying hadron production in hadron-proton, hadron-nucleus and nucleus-nucleus collisions at the CERN SPS. The programme on strong interactions includes the study of the properties of the onset of deconfinement \cite{pedestrians} and the search for the critical point of strongly interacting matter \cite{proposal}. Additionally two large projects of reference measurements are ongoing on requests of neutrino and cosmic ray physics.


The new results obtained within the strong interaction programme, that we highlight here, concern transverse mass spectra and mean multiplicities of the most abundant charged hadrons ($\pi^{-}, \pi^{+}, K^-$ and $K^+$) produced in central collisions of intermediate size ions, $^{7}$Be+$^{9}$Be \cite{BeBe} and $^{40}$Ar+$^{45}$Sc \cite{ArSc}, in the beam momentum range of 19\textit{A}--150\textit{A} GeV/\textit{c} ($\sqrt{s_{NN}}=5.12$--$16.83$ GeV). The particle identification was performed using data on specific energy loss ($dE/dx$) measured by four large volume TPCs, which was supplemented by the measurement of time of flight ($tof$) in a region where $dE/dx$ distribution for various particle species overlap (low momenta $p\lesssim7$ GeV/\textit{c}). This yields results in a large acceptance --- NA61/SHINE measures particles down to zero transverse momentum in almost complete forward hemisphere.
Consequently mean multiplicities of hadrons produced in whole phase-space can be extracted. These are very important traits of fixed-target experiments.

A proper selection of most central events presents a unique experimental challenge, especially so in case of collisions of small ions.
While proton-proton interactions can be classified as elastic or inelastic and central collisions of very large ions of lead can be in some sense distinguished based on the multiplicity of produced particles, neither method is well suited for "intermediate" systems like Be+Be or Ar+Sc.
Moreover, the latter method may lead to a bias, as particle multiplicity depends on physics of interest.
A procedure devised by NA61/SHINE relies on the measurement of forward energy $E_F$ of collision spectators in a modular calorimeter called Projectile Spectator Detector.
The most central collisions deposit the smallest energy $E_F$. This allows for a precise, reproducible and, what is crucial, unbiased selection of central events.
Measurement of forward energy of projectile spectators is yet another trait reserved for fixed-target experiments.

\section{Small and large systems}

The comparative study of collisions of protons and heavy ions offer an incredibly rich source of information of collective behavior of strongly interacting matter. It enables us to reach beyond elementary interactions and ask more involved questions: what are the phases of strongly interacting matter? How do the transitions between these phases happen? After almost 50 years of measurements of hadron production at broad range of collisions energies and countless theoretical attempts of explanation of quarks and gluons interplay we have some answers, but as it is usually the case in physics --- even more questions.

In the early 2000's the discovery of the new, deconfined state of strongly interacting matter created in heavy ion collisions was announced by CERN \cite{CERN_QGP} and RHIC \cite{Adcox:2001jp}. 
Just a couple years later the NA49 experiment identified the energy at which the deconfinement phase-transition takes place \cite{NA49_OOD}. The Collaboration published the measurements of Pb+Pb interactions at five collision energies \cite{NA49_PbPb, NA49_OOD}, showing rapid changes of collision energy dependence of basic hadron production properties, among them:
\begin{itemize}
    \item \textbf{step} --- a plateau in the inverse slope parameter of transverse mass spectra of kaons,
    \item \textbf{horn} --- a sharp peak in the $K^+/\pi^+$ ratio.
\end{itemize}
These signatures were in agreement with the first model implementing deconfinement in heavy-ion collisions, the Statistical Model of Early Stage \cite{SMES}. Moreover, they both appeared at the same collision energy (see: Pb+Pb and Au+Au data in Figs. \ref{fig:horn} and \ref{fig:step}). NA49's results were therefore interpreted as an evidence of the \textit{onset of deconfinement} --- an energy at which the matter created at the early stage of the collision appears in the form of quark-gluon plasma.

Phase structure of hadronic matter becomes more and more involved correspondingly to progresses in theoretical understanding of the subject and collecting more and more experimental data. While the largest experimentally available now energies at LHC and RHIC colliders seem to provide data related to the crossover region between quark-gluon plasma and hadron gas then the SPS fixed-target NA61/SHINE experiment is particularly suited to explore the hypothetical first-order phase transition line with the critical point included.


The phases of QCD matter depend on two primary parameters: temperature $T$ and baryo-chemical potential $\mu_B$.
With changing the energy of the collision we can traverse the phase-diagram in a particular direction --- increasing collision energies results in smaller $\mu_B$ and larger $T$.
In order to study a broader domain of the QCD phase-diagram we also vary the sizes of colliding nuclei.
Hence with the NA61/SHINE's unique two-dimensional scan in collision energy and system size we seek the signatures of the onset of deconfinement, aiming to pinpoint the location of the phase-transition line.

\section{Intermediate size ions}

\begin{figure}[h]
    \centering
    \includegraphics[width=0.48\textwidth]{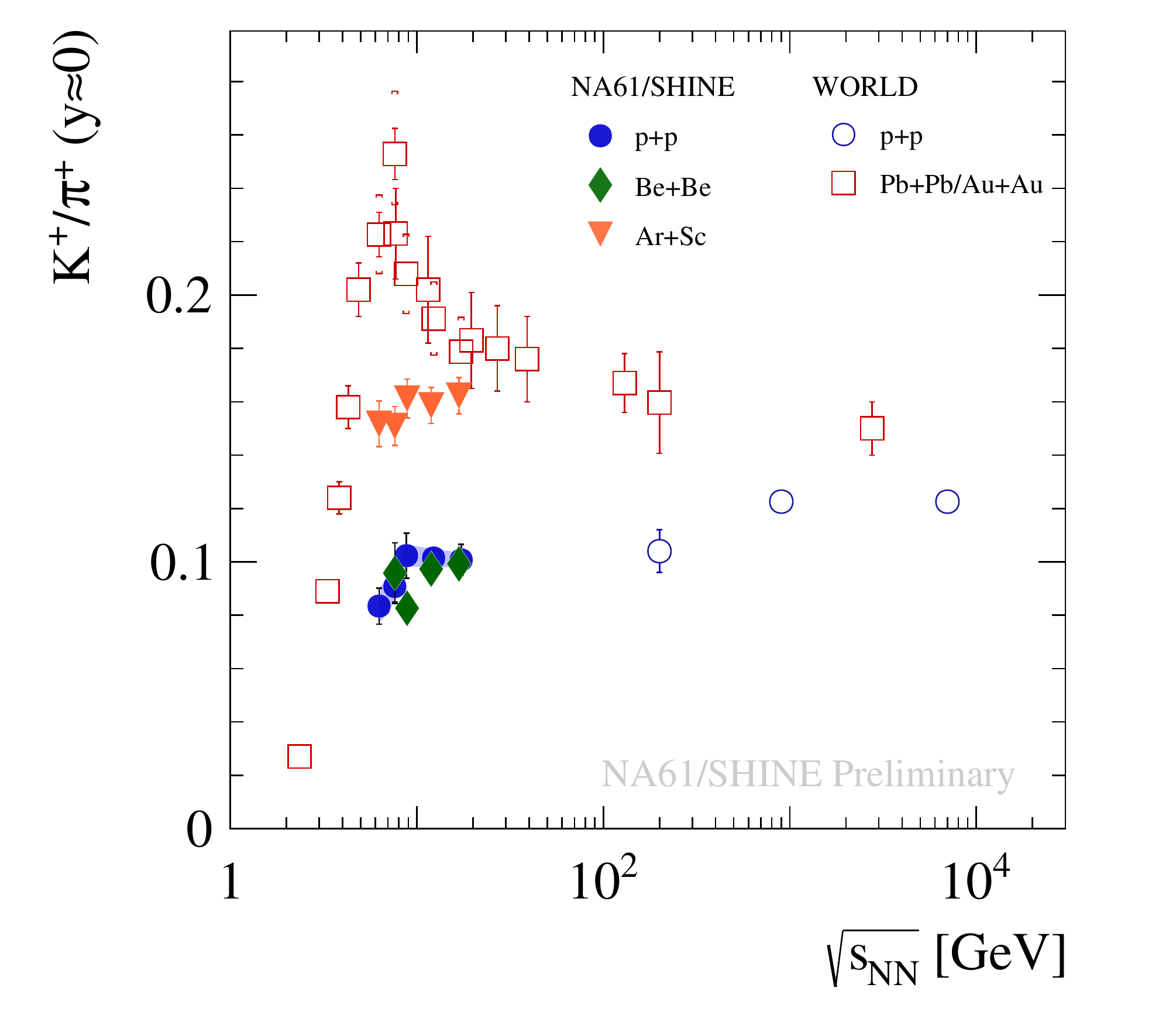}
    \includegraphics[width=0.48\textwidth]{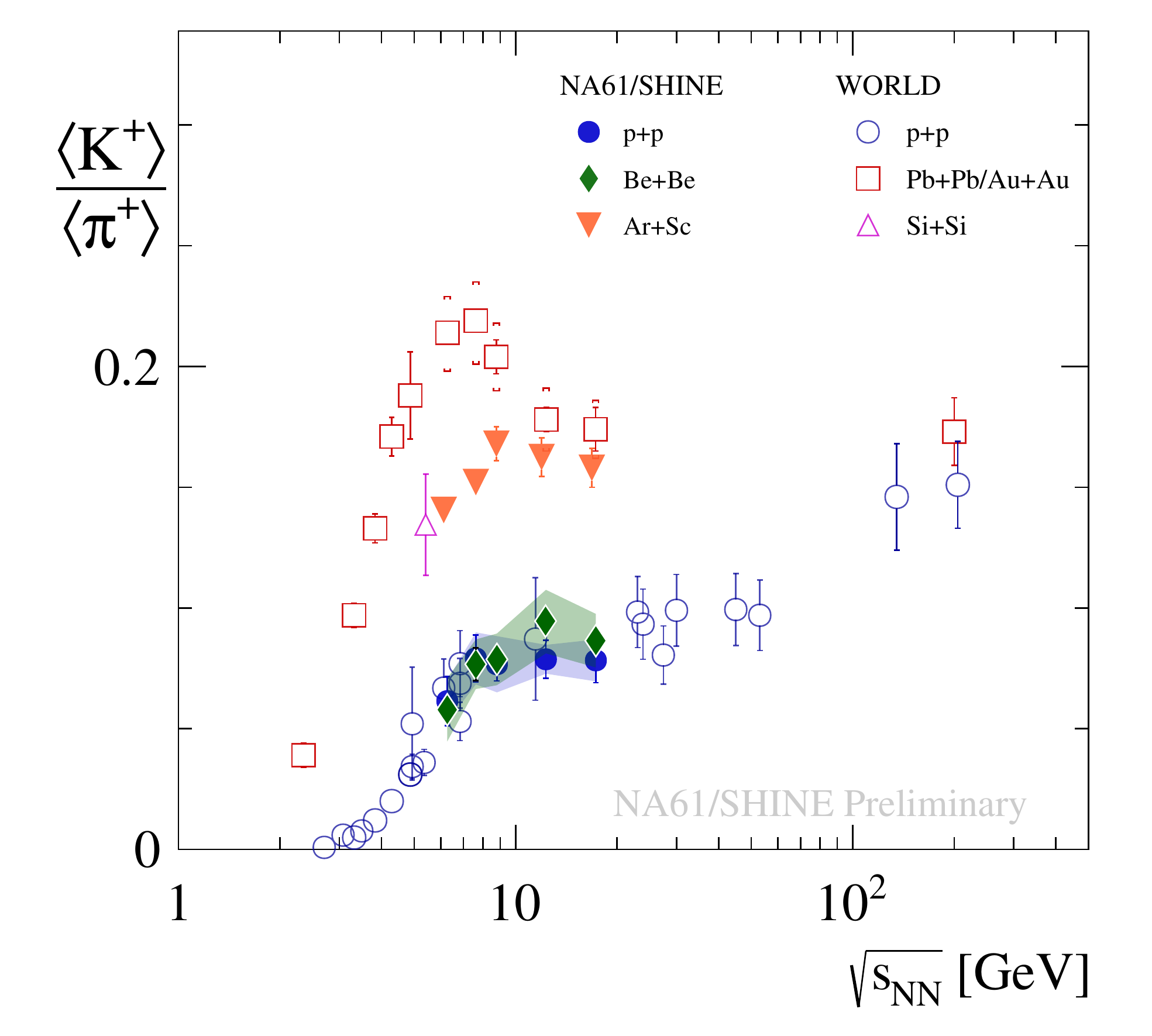}
    \caption{Measurements of the $K^+/\pi^+$ ratio at mid-rapidity (left plot) and in whole phase space "$4\pi$" (on the right) in dependence on collision energy $\sqrt{s_{NN}}$. Very pronounced differences between data on p+p (blue) and Pb+Pb (red) collisions were identified as a signature of the onset of deconfinement in Pb+Pb interactions. New data on intermediate size systems, collisions of Be+Be (green) and Ar+Sc (orange), were recently measured by NA61/SHINE.}
    \label{fig:horn}
\end{figure}

So what do we learn from the new NA61/SHINE's data on Be+Be and Ar+Sc reactions?
Let us focus on the most prominent signature of the onset of deconfinement --- the "horn".
The phase transition seems apparent in the characteristic, non-monotonic behavior of the kaon over pion ratio in central heavy ion collisions (see: Pb+Pb and Au+Au in Fig. \ref{fig:horn}).
In case of intermediate size systems however, no such vivid structure is present. Nevertheless, two interesting features demand attention. Firstly, a clear distinction between two data sets is visible --- p+p and Be+Be results cluster around similar values, while Pb+Pb, Au+Au and Ar+Sc show much higher $K^+/\pi^+$ ratios. Secondly, although Ar+Sc is clearly separated from small systems its energy dependence does not resemble the sharp peak seen in heavy-ion reactions \cite{Aduszkiewicz:2287091}.
No theoretical description can reproduce this behavior --- neither statistical \cite{SMES, motornenko} nor dynamical models \cite{PHSD}.

\begin{figure}[h]
    \centering
    \includegraphics[width=0.48\textwidth]{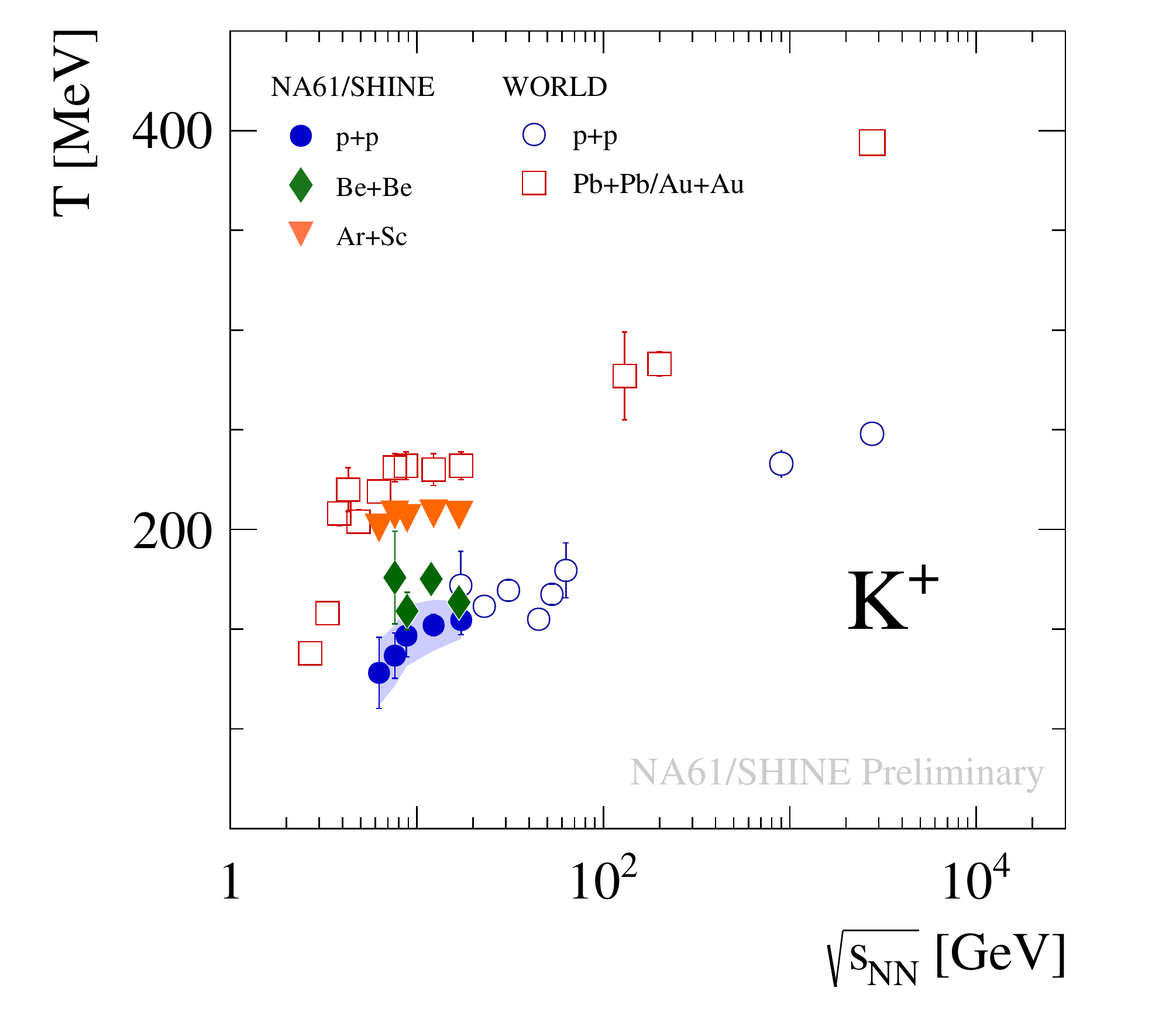}
    \includegraphics[width=0.48\textwidth]{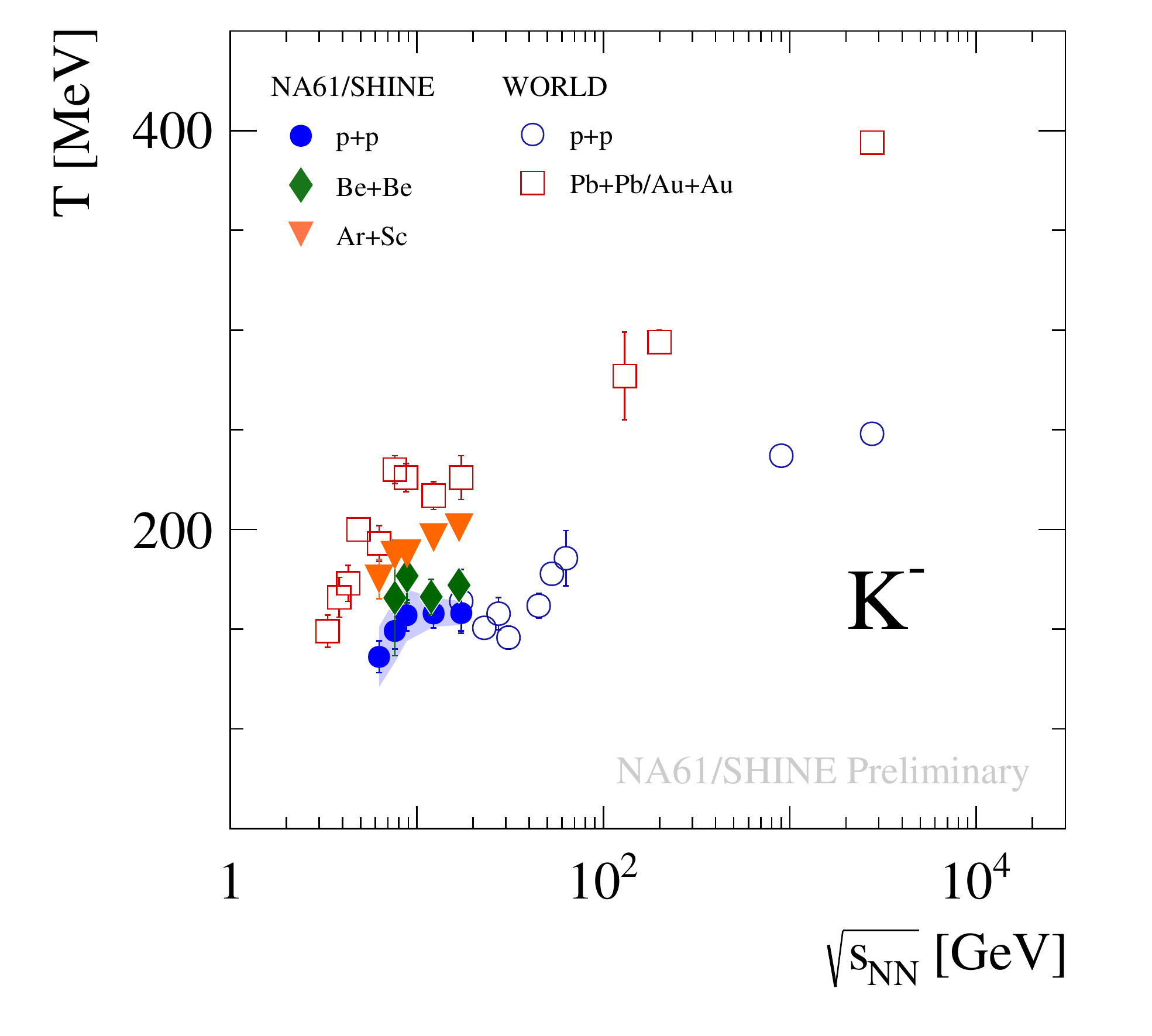}
    \caption{Collision energy dependence of the inverse slope parameter $T$ of transverse momentum spectra of charged kaons, $K^+$ and $K^-$. Note a plateau visible in measurements of Pb+Pb collisons at $\sqrt{s_{NN}}$ $\approx$ $10$ GeV/\textit{c} which is interpreted as a signature of the onset of deconfinement. Similar structure is also visible for smaller systems: Be+Be and Ar+Sc.}
    \label{fig:step}
\end{figure}

Complementary to the above, Fig. \ref{fig:step} displays the energy dependence of the inverse slope $T$, fitted to the transverse mass spectra of charged kaons ($K^+$ and $K^-$). In central heavy ion collisions we see a characteristic step-like behavior. Under assumption that the parameter $T$ can be treated as a vague analogy of temperature, it can be considered as a strong interactions analogy of the ice-water phase-transition. Once again we notice the proximity of p+p and Be+Be measurements. Similarly $T$ obtained for Ar+Sc collisions is higher, but not as high as for heavy lead nuclei.

\section{Onset of fireball}
\begin{figure}
    \centering
    \includegraphics[width=0.6\textwidth]{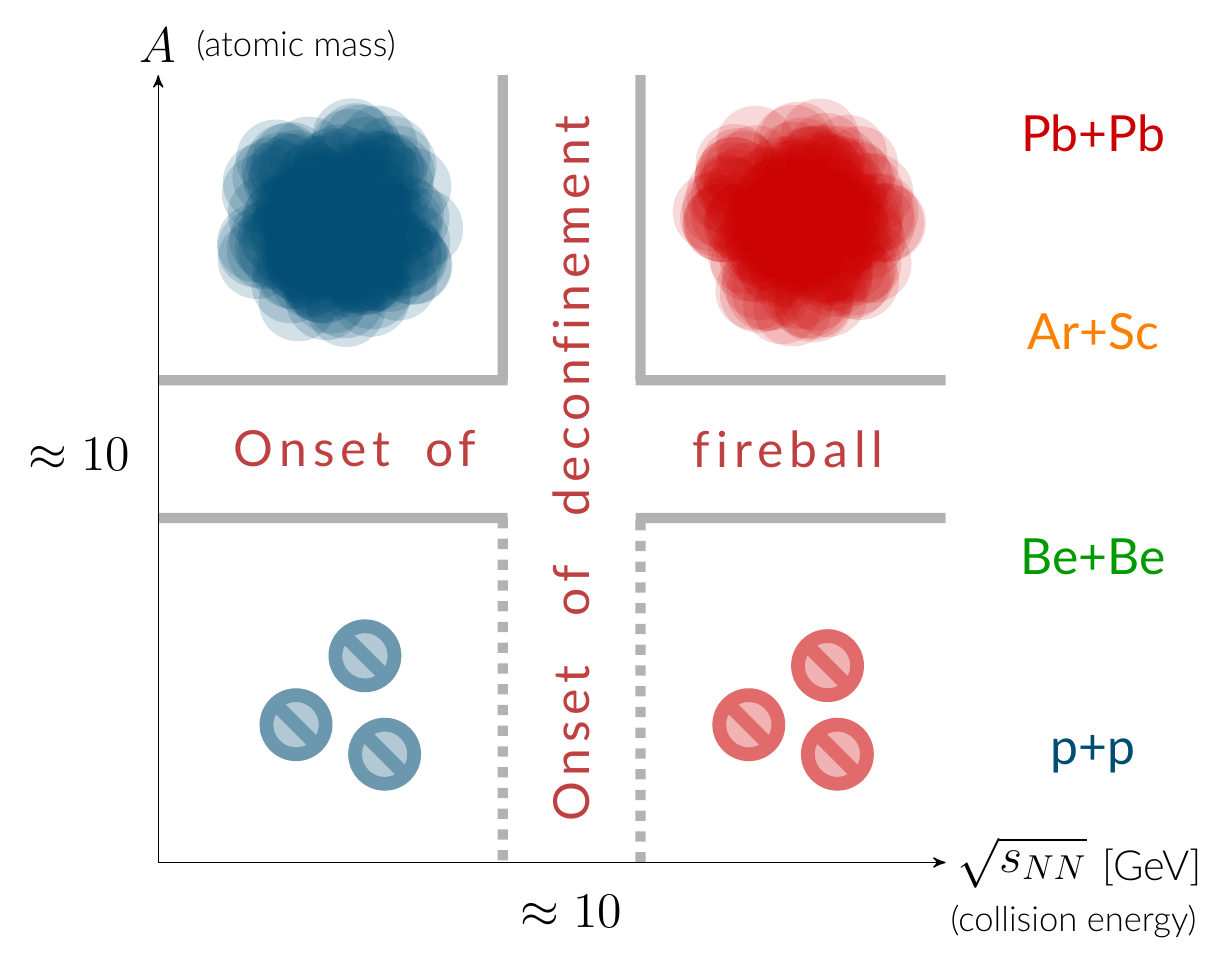}
    \caption{The diagram shows hypothesized four domains that can be assigned to collisions in the SPS energy range. On top of the onset of deconfinement, there is an onset of fireball -- beginning of creation of large clusters with increasing system size. It is still unclear what happens with collisions of small nuclei with increasing collision energy. }
    \label{fig:domains}
\end{figure}

The observed rapid change of hadron production properties that starts when moving from Be+Be to Ar+Sc collisions hints a beginning of the creation of large clusters of strongly interacting matter --- the \textit{onset of fireball} \cite{Larsen:2018oon}. The similarities of p+p and Be+Be systems suggests that interactions of these systems could form small non-equilibrium clusters via binary collisions of nucleons, exactly like in the Wounded Nucleon Model \cite{WNM}.
On the other hand properties of Pb+Pb collisions are pretty well described by statistical and hydrodynamical models, which assume creation of collectively evolving fireball.
Results on Ar+Sc collisions are clearly closer to the Pb+Pb ones than to p+p and Be+Be measurements.
Summarizing, we could distinguish four domains of collision dynamics, split by the two critical phenomena: the onset of deconfinement and the onset of fireball, as seen in Fig. \ref{fig:domains}.
The onset of deconfinement from confined matter to quark-gluon plasma is well established in the fireball region at high masses of the colliding nuclei, while its presence in the small cluster regime is still an open question --- an incredibly interesting question, certainly deserving a separate study.

\section{Extension beyond 2020}
The quest for understanding the phases of strongly interacting matter continues and many unanswered questions remain \cite{Turko:2018hty}. In the following we briefly present the next steps of NA61/SHINE.

The second stage of the NA61 experiment, starting after the Long Shutdown 2 (LS2) would include \textbf{measurements of charm hadron production} in Pb+Pb collisions.
The main objective is to obtain the first data on mean number of $\bar{c}c$ pairs produced in the full phase space in heavy ion collisions.
To highlight the need for such measurements let us note that the theoretical predictions concerning $\langle c\bar{c} \rangle$ production at top SPS energy differ by about two orders of magnitude \cite{Larsen:2019due}.  Moreover, further new results will give us new insights on the collision energy and centrality dependence. These data will help to answer the long-standing questions about the mechanism of open charm production, about the relation between the onset of deconfinement and open charm production, and about the behaviour of $J/\psi $ in quark-gluon plasma.


NA61/SHINE is the only experiment which will conduct such measurements in the near future. Together with other heavy ion experiments it creates a full-tone physical picture of QCD in dense medium.



\begin{thebibliography}{99}
\bibitem{facility}
  N.~Abgrall {\it et al.} [NA61/SHINE Collaboration],
  JINST {\bf 9} (2014) P06005

\bibitem{pedestrians}
  M.~Gazdzicki, M.~Gorenstein and P.~Seyboth,
  Acta Phys.\ Polon.\ B {\bf 42} (2011) 307

\bibitem{proposal}
  N.~Antoniou {\it et al.} [NA49-future Collaboration],
  CERN-SPSC-P-330.
  \textit{(NA61/SHINE Proposal)}
  
\bibitem{BeBe}
  S.~Pulawski [for NA61/SHINE Collaboration],
  arXiv:1911.01398 [nucl-ex]

\bibitem{ArSc}
  P. Podlaski [for NA61/SHINE Collaboration],
  SQM Bari 2019, \url{sqm2019.ba.infn.it/}
  
\bibitem{Pelaez:2015qba}
  J.~R.~Pelaez,
  Phys.\ Rept.\  {\bf 658}, 1 (2016)

\bibitem{CERN_QGP}
  CERN Press Release, 10 FEBRUARY, 2000, 
  \url{home.cern/news/press-release/cern/new-state-matter-created-cern},
  
  \bibitem{Adcox:2001jp} 
  K.~Adcox {\it et al.} [PHENIX Collaboration],
  Phys.\ Rev.\ Lett.\  {\bf 88}, 022301 (2002)
  doi:10.1103/PhysRevLett.88.022301
  [nucl-ex/0109003].
  
\bibitem{NA49_OOD}
  C.~Alt {\it et al.} [NA49 Collaboration],
  Phys.\ Rev.\ C {\bf 77} (2008) 024903
  
\bibitem{NA49_PbPb}
  S.~V.~Afanasiev {\it et al.} [NA49 Collaboration],
  Phys.\ Rev.\ C {\bf 66} (2002) 054902

\bibitem{SMES}
  M.~Gazdzicki and M.~I.~Gorenstein,
  Acta Phys.\ Polon.\ B {\bf 30} (1999) 2705


\bibitem{motornenko}
  A.~Motornenko, V.~V.~Begun, V.~Vovchenko, M.~I.~Gorenstein and H.~Stoecker,
  Phys.\ Rev.\ C {\bf 99} (2019) no.3,  034909
  , and references therein
  
\bibitem{PHSD}
  A.~Palmese, W.~Cassing, E.~Seifert, T.~Steinert, P.~Moreau and E.~L.~Bratkovskaya,
  Phys.\ Rev.\ C {\bf 94} (2016) no.4,  044912

\bibitem{Aduszkiewicz:2287091}
  Aduszkiewicz, A. [NA61/SHINE Collaboration]
  Technical Report SPSC-SR-221







  
\bibitem{WNM}
  A. Bia{\l}as, M. Bleszy{\'n}ski, and W. Czy{\.z}
  Nucl.\ Phys.\ B {\bf 111} (1976) 461

\bibitem{Larsen:2018oon}
  Larsen, D. [for NA61/SHINE Collaboration]
  EPJ Web Conf.\  {\bf 199}, 04010 (2019)
  
 \bibitem{Turko:2018hty} 
  L.~Turko [for NA61/SHINE Collaboration],
  Particles {\bf 1}, 296 (2018)
  
\bibitem{Larsen:2019due} 
  D.~Larsen [NA61/SHINE Collaboration),
  Universe {\bf 5}, 24 (2019)


\end{thebibliography}
\end{document}